\documentclass[aps, prd, superscriptaddress, twocolumn]{revtex4-1}

\usepackage{wallpaper}
\usepackage{bm}
\usepackage{amsmath}
\usepackage[colorlinks,linkcolor=blue,anchorcolor=blue,citecolor=red]{hyperref}

\begin{document}

\title{Strangelets at finite temperature: nucleon emission rates, interface and shell effects}

\author{Hao-Song~You}
\affiliation{Center for Gravitation and Cosmology, College of Physical Science and Technology, Yangzhou University, Yangzhou 225009, China}

\author{Huai-Min~Chen}
\affiliation{School of Mechanical and Electrical Engineering, Wuyi University, Wuyishan 354300, China}

\author{Jian-Feng~Xu}
\affiliation{College of information Engineering, Suqian University, Suqian 223800, China}

\author{Cheng-Jun~Xia}
\email{cjxia@yzu.edu.cn}
\affiliation{Center for Gravitation and Cosmology, College of Physical Science and Technology, Yangzhou University, Yangzhou 225009, China}

\author{Guang-Xiong~Peng}
\affiliation{School of Nuclear Science and Technology, University of Chinese Academy of Sciences, Beijing 100049, China}
\affiliation{Theoretical Physics Center for Science Facilities, Institute of High Energy Physics, Beijing 100049, China}
\affiliation{Synergetic Innovation Center for Quantum Effects and Application, Hunan Normal University, Changsha 410081, China}

\author{Ren-Xin~Xu}
\affiliation{School of Physics, Peking University, Beijing 100871, China}
\affiliation{Kavli Institute for Astronomy and Astrophysics, Peking University, Beijing 100871, China}

\date{\today}

\begin{abstract}
We investigate the properties of strangelets at finite temperature $T$, where an equivparticle model is adopted with both the linear confinement and leading-order perturbative interactions accounted for using density-dependent quark masses. The shell effects are examined by solving the Dirac equations for quarks in the mean-field approximation, which diminish with temperature as the occupation probability of each single-particle levels fixed by the Fermi-Dirac statistics, i.e., shell dampening. Consequently, instead of decreasing with temperature, the surface tension extracted from a liquid-drop formula increases with $T$ until reaching its peak at $T\approx 20$-40 MeV with vanishing shell corrections, where the formula roughly reproduces the free energy per baryon of all strangelets. The curvature term, nevertheless, decreases with $T$ despite the presence of shell effects. The neutron and proton emission rates are fixed microscopically according to the external nucleon gas densities that are in equilibrium with strangelets, which generally increase with $T$ ($\lesssim 50$ MeV) for stable strangelets but decrease for those that are unstable against nucleon emission at $T=0$. The energy, free energy, entropy, charge-to-mass ratio, strangeness per baryon, and root-mean-square radius of $\beta$-stable strangelets obtained with various parameter sets are presented as well. The results indicated in this work are useful for understanding the products of binary compact star mergers and heavy-ion collisions.
\end{abstract}


\maketitle

\section{\label{sec:intro}Introduction}

At large enough densities, hadronic matter will undergo a deconfinement phase transition and form quark matter~\cite{Fukushima2005_PRD71-034002, Voskresensky2023_PPNP130-104030}. In 1971, Bodmer proposed the possible existence of ``collapsed nuclei" with mass number $A > 1$ and radii much smaller than ordinary nuclei~\cite{Bodmer1971_PRD4-1601}. Later on, Witten suggested that strange quark matter (SQM) carrying roughly equal numbers of $u$, $d$, and $s$ quarks and a small amount of electrons may be the ground state of strongly interacting system~\cite{Witten1984_PRD30-272}. If true, there must exist various types of SQM objects, e.g., strangelets~\cite{Farhi1984_PRD30-2379, Berger1987_PRC35-213, Gilson1993_PRL71-332, Peng2006_PLB633-314}, nuclearites~\cite{Rujula1984_Nature312-734, Lowder1991_NPB24-177}, meteorlike compact ultradense objects~\cite{Rafelski2013_PRL110-111102}, and strange stars~\cite{Itoh1970_PTP44-291, Alcock1986_ApJ310-261, Haensel1986_AA160-121}. Nevertheless, chiral models suggest that SQM is unstable~\cite{Buballa1999_PLB457-261, Klahn2015_ApJ810-134} and nonstrange quark matter ($ud$QM) may be the true ground state~\cite{Holdom2018_PRL120-222001}, indicating the possible existence of stable $ud$QM nuggets with $A\gtrsim 300$~\cite{Holdom2018_PRL120-222001} and nonstrange quark stars~\cite{Zhao2019_PRD100-043018, Zhang2020_PRD101-043003, Cao2020, Zhang2021_PRD103-063018, Wang2021_Galaxies9-70, Xia2022_PRD106-034016, Yuan2022_PRD105-123004, Restrepo2023_PRD107-114015}. Additionally, instead of deconfined quark matter, a solid state comprised of strangeons (quark-clusters with three-light-flavor symmetry) may be more stable~\cite{Xu2003_ApJ596-L59, Lai2017_JPCS861-012027, Xu2018_SCPMA61-109531, Miao2022_IJMPE0-2250037, Zhang2023_PRD108-063002}, where small strangeon nuggets could persist in our universe~\cite{Xu2019_AIPCP2127-020014}.
Such exotic objects are expected to be formed in heavy-ion collisions~\cite{Weiner2006_IJMPE15-37}, the mergers of binary compact stars~\cite{Madsen2002_JPG28-1737, Madsen2005_PRD71-014026, Bauswein2009_PRL103-011101, Lai2018_RAA18-024, Lai2021_RAA21-250, Bucciantini2022_PRD106-103032}, type~\uppercase\expandafter{\romannumeral2} supernova explosions~\cite{Vucetich1998_PRD57-5959}, the hadronization process of the early universe~\cite{Witten1984_PRD30-272}, etc. Significant efforts were then devoted to search for those objects, but with no conclusive evidence~\cite{Finch2006_JPG32-S251, Burdin2015_PR582-1}. It is thus essential for us to disentangle the properties of those exotic objects to finally prove or disprove their existence.

As lattice QCD is plagued by the sign problem at finite densities, we rely on effective models to unveil the properties of those exotic objects. For example, adopting MIT bag model, Farhi and Jaffe have found that the surface tension has a significant impact on the stability of strangelets~\cite{Farhi1984_PRD30-2379}. Later on, Berger and Jaffe proposed a mass formula for strangelets and investigated their possible decay channels~\cite{Berger1987_PRC35-213}. Small strangelets are almost uniformly charged, while the charge screening effects start to play a role for larger strangelets~\cite{Heiselberg1993_PRD48-1418}. In particular, when the surface tension is below some critical value, the surface of a quark star will fragment into a crystalline crust made of charged strangelets or $ud$QM nuggets immersed in an electron gas~\cite{Alford2006_PRD73-114016, Jaikumar2006_PRL96-041101, Xia2022_PRD106-034016}. For small strangelets and $ud$QM nuggets, it was shown that the curvature contribution is sizable~\cite{Lugones2021_PRC103-035813}, where the multiple reflection expansion method (MRE) was proposed to treat the surface and curvature corrections analytically~\cite{Madsen1993_PRL70-391, Madsen1993_PRD47-5156, Madsen1994_PRD50-3328}. The shell effects in strangelets was addressed as well in the framework of MIT bag model~\cite{Terazaw1989_JPSJ58-3555, Terazaw1989_JPSJ58-4388, Terazaw1990_JPSJ59-1199, Madsen1994_PRD50-3328, Oertel2008_PRD77-074015}, which alters the properties of small strangelets significantly.

Those investigations on strangelets and $ud$QM nuggets are generally carried out at vanishing temperatures, while their creation and survival usually take place at large temperatures. It is thus necessary to investigate the properties of those objects at finite temperatures. This was done within the framework of MIT bag model, where clear shell structures persist up to about $T=10$ MeV were identified~\cite{Mustafa1996_PRD53-5136, Mustafa1997_PRC56-420}. For quark nuggets in a color-flavor locked state, a substantial quenching of the evaporation and boiling processes was identified in the cosmological quark-hadron transition~\cite{Lugones2004_PRD69-063509}. Nevertheless, it is worth mentioning that MIT bag model assumes an infinite wall while lattice QCD suggests linear confinement for quarks, leading to distinctive different surface density profiles and consequently affecting various strangelets' properties as indicated in our previous study~\cite{Xia2018_PRD98-034031}. We thus adopt an equivparticle model here to study the properties of strangelets, where both the linear confinement and leading-order perturbative interactions can be accounted for using density-dependent quark masses~\cite{Peng2000_PRC62-025801, Wen2005_PRC72-015204, Wen2007_JPG34-1697, Chen2012_CPC36-947, Xia2014_PRD89-105027}. Note that the properties of strangelets at finite temperatures were investigated in the framework of equivparticle model with the interface effects treated with the MRE method~\cite{Zhang2003_PRC67_015202, Wen2005_PRC72-015204, Chen2022_PRD105-014011}, where the mass, radius, and strangeness per baryon are increasing with temperature and charge-to-mass ratio decreasing with temperature~\cite{Chen2022_PRD105-014011}.

In this work, as was done in our previous study~\cite{Xia2018_PRD98-034031}, we investigate the properties of strangelets in mean-field approximation (MFA). Instead of adopting MRE method, we directly solve the Dirac equations and obtain the single particle levels for quarks. The calculation is carried out self-consistently with the mean fields and quark wave functions obtained in an iterative manner, where the occupation probability of each single particle level follows Fermi-Dirac distribution at given temperature. Then we fix the neutron and proton emission rates according to the densities of external neutron and proton gases in equilibrium with the strangelet~\cite{Zhu2014_PRC90-054316}.

This paper is organized as follows. In Sec.~\ref{sec:the} we present the theoretical framework of our study, where the Lagrange density of equivparticle model with density dependent quark masses is given in Sec.~\ref{sec:the_Lagrangian}, the properties of strangelets in MFA are obtained in Sec.~\ref{sec:the_SletSM}, and neutron and proton emission rates are fixed in Sec.~\ref{sec:Emission rates}. The numeric result on the properties strangelets at finite temperature are presented in Sec.~\ref{sec:num}. Finally, our conclusion is given in Sec.~\ref{sec:con}.

\section{\label{sec:the}Theoretical framework}
\subsection{\label{sec:the_Lagrangian}Lagrangian density}

The Lagrangian density of equivparticle model can be given as
\begin{equation}
\mathcal{L} =  \sum_{i=u,d,s} \bar{\psi}_i \left[ i \gamma^\mu \partial_\mu - m_i - e q_i \gamma^\mu A_\mu \right]\psi_i - \frac{1}{4} A_{\mu\nu}A^{\mu\nu},  \label{eq:Lgrg_all}
\end{equation}
where $\psi_i$ represents the Dirac spinor of quark flavor $i$, $m_i$ the mass, $q_i$ the charge ($q_u = 2e/3$ and $q_d = q_s = -e/3$), and $A_\mu$ the photon field with the field tensor
\begin{equation}
A_{\mu\nu} = \partial_\mu A_\nu - \partial_\nu A_\mu.
\end{equation}
In the equivparticle model, the strong interactions among quarks are treated with density and/or temperature-dependent quark masses, where quarks can be regarded as quasi-free particles. The mass of quark $i$ is usually fixed by
\begin{equation}
m_i = m_{i0} + m_\mathrm{I}(n_{\mathrm{b}},T),
\end{equation}
where $n_{\mathrm{b}} = \sum_{i=u,d,s} {n_i}/{3}$ is the baryon number density with $n_i=\langle {\bar{\psi}}_i\gamma^0\psi_i\rangle$ being the number density of quark $i$, $T$ the temperature, and $m_{u0}=2.2$ MeV, $m_{d0}=4.7$ MeV, and $m_{s0}=96.0$ MeV the current quark masses~\cite{PDG2016_CPC40-100001}. The original quark mass scaling was derived from MIT bag model~\cite{Fowler1981_ZPC9-271, Chakrabarty1989_PLB229-112, Benvenuto1995_PRD51-1989}, which is given by
\begin{equation}
m_\mathrm{I} = \frac{B}{3n_{\mathrm{b}}}
\end{equation}
with $B$ being the bag constant. Considering the contributions of linear confinement and in-medium chiral condensates, an inversely cubic scaling was proposed~\cite{Peng1999_PRC61-015201}, i.e.,
\begin{equation}
m_\mathrm{I} = \frac{D}{\sqrt[3]{n_\mathrm{b}}}, \label{eq:mI_D}
\end{equation}
where the confinement parameter $D$ is connected to the string tension of linear confinement and vacuum chiral condensates. By adding one more term to Eq.~(\ref{eq:mI_D}), the one-gluon-exchange interaction or lead-order perturbation interaction can be considered~\cite{Chen2012_CPC36-947, Xia2014_PRD89-105027}, i.e.,
\begin{equation}
m_\mathrm{I} = \frac{D}{\sqrt[3]{n_\mathrm{b}}}+C \sqrt[3]{n_\mathrm{b}}. \label{eq:mI_CD}
\end{equation}
Depending on the sign of $C$, the second term corresponds to the contribution from one-gluon-exchange interaction ($C<0$)~\cite{Chen2012_CPC36-947} or leading-order perturbative interaction ($C>0$)~\cite{Xia2014_PRD89-105027}. To accommodate the deconfinement phase transition at high temperatures, this mass scaling was later extended and became temperature dependent~\cite{Lu2016_NST27-148}, i.e.,
\begin{equation}
m_\mathrm{I} = \frac{D}{\sqrt[3]{n_\mathrm{b}}}\left(1+\frac{8T}{\Lambda}e^{-\frac{\Lambda}{T}}\right)^{-1}+C\sqrt[3]{n_\mathrm{b}}\left(1+\frac{8T}{\Lambda}e^{-\frac{\Lambda}{T}}\right),
\label{eq:mI_CDL}
\end{equation}
where $\Lambda=280$ MeV is a temperature scale parameter corresponding to the critical temperature $T_\mathrm{c}\approx 175$ MeV~\cite{Wen2005_PRC72-015204}. As strangelets will not exist at large temperatures at $T \gtrsim T_\mathrm{c}$ and the variation of $m_\mathrm{I}$ with respect to temperature is generally small at $T \lesssim T_\mathrm{c}$, we thus neglect the temperature dependence of quark masses and simply adopt Eq.~(\ref{eq:mI_CD}) in this work.

\subsection{\label{sec:the_SletSM} Strangelets in MFA}
Based on the Lagrangian density indicated in Eq.~(\ref{eq:Lgrg_all}) with the quark mass scaling in Eq.~(\ref{eq:mI_CD}), the Dirac equations for quarks and Klein-Gordon equation for photons are obtained via a standard variational procedure. For spherically symmetric strangelets, the Dirac spinor of quarks can be expanded as
\begin{equation}
 \psi_{n\kappa m}({\bm r}) =\frac{1}{r}
 \left(\begin{array}{c}
   iG_{n\kappa}(r) \\
    F_{n\kappa}(r) {\bm\sigma}\cdot{\hat{\bm r}} \\
 \end{array}\right) Y_{jm}^l(\theta,\phi)\:,
\label{EQ:RWF}
\end{equation}
where $G_{n\kappa}(r)/r$ and $ F_{n\kappa}(r)/r$ are the radial wave functions for the upper and lower components, and $Y_{jm}^l(\theta,\phi)$ the spinor spherical harmonics. The quantum number $\kappa$ is defined by the angular momenta $(l, j)$ as $\kappa=(-1)^{j+l+1/2}(j+1/2)$ with $j = l \pm \frac{1}{2}$.

Adopting mean-field approximation and substituting Eq.~\eqref{EQ:RWF} into Dirac equation, the one-dimensional radial Dirac equation can then be easily obtained by integrating the angular part, i.e.,
\begin{equation}
 \left(\begin{array}{cc}
  V_{iV} + V_{iS}                                               & {\displaystyle -\frac{\mbox{d}}{\mbox{d}r} + \frac{\kappa}{r}}\\
  {\displaystyle \frac{\mbox{d}}{\mbox{d}r}+\frac{\kappa}{r}} & V_{iV} - V_{iS}                       \\
 \end{array}\right)
 \left(\begin{array}{c}
  G_{in\kappa} \\
  F_{in\kappa} \\
 \end{array}\right)
 = \varepsilon_{in\kappa}
 \left(\begin{array}{c}
  G_{in\kappa} \\
  F_{in\kappa} \\
 \end{array}\right) \:,
\label{Eq:RDirac}
\end{equation}
Here $\varepsilon_{in\kappa}$ is the single particle energy of quark $i$, the mean-field scalar and vector potentials of quarks are as follows
\begin{eqnarray}
 V_{iS} &=& m_{i0} + m_\mathrm{I}(n_\mathrm{b}), \label{Eq:Vs}\\
 V_{iV} &=& V_V + e q_i A_0. \label{Eq:Vv}
\end{eqnarray}
Note that we have added current mass of quarks to the scalar potential $V_S = m_\mathrm{I}(n_\mathrm{b})$ in Eq.~\eqref{Eq:Vs}. In the vector potentials we have $V_V = \frac{1}{3}\frac{\mbox{d} m_\mathrm{I}}{\mbox{d} n_\mathrm{b}}\sum_{i=u,d,s}  n_i^\mathrm{s}$, which arises due to the density-dependence of quark masses and is essential to meet the requirement of thermodynamic self-consistency~\cite{Xia2014_PRD89-105027, Xia2018_PRD98-034031}. The Klein-Gordon equation for photons is given by
\begin{equation}
- \nabla^2 A_0 = e n_\mathrm{ch}. \label{Eq:K-G}
\end{equation}
where $n_\mathrm{ch} = \sum_i q_i n_i$ is the charge density with $q_u = 2/3, q_d = -1/3$, and $q_s = -1/3$.

Once we fix the single particle energies for quarks, their occupation probability is then determined by the Fermi-Dirac distribution, i.e.,
\begin{equation}
f_{in\kappa} = \left[1 + e^{( \varepsilon_{in\kappa} - \mu_i)/T}\right]^{-1},  \label{Eq:F-D}
\end{equation}
where $\mu_i$ represents the chemical potential of quark favor $i$ and we consider only the $\beta$-equilibrated cases with \begin{equation}
\mu_u=\mu_d=\mu_s=\mu_\mathrm{b}/3. \label{Eq:bstb}
\end{equation}
Note that we have adopted the no-sea approximation and neglected any contributions from anti-quarks, which is justified here since the adopted temperatures are small. As neutrinos can leave the system freely and electrons has little impact on the properties of strangelets with radii $R\lesssim 40$ fm~\cite{Xia2017_JPCS861-012022}, the contribution of neutrinos and electrons are neglected by taking their chemical potentials equal to zero. The scalar and vector densities for quarks at finite temperatures can then be obtained with
\begin{subequations}
\begin{eqnarray}
 n_i^\mathrm{s}(r) &=& \sum_{n, \kappa} \frac{3|\kappa| f_{in\kappa}}{2\pi r^2}
 \left[|G_{in\kappa}(r)|^2-|F_{in\kappa}(r)|^2\right] \:,
\\
 n_i(r) &=& \sum_{n, \kappa} \frac{3|\kappa| f_{in\kappa}}{2\pi r^2}
 \left[|G_{in\kappa}(r)|^2+|F_{in\kappa}(r)|^2\right] \:,
\end{eqnarray}%
\label{Eq:Density}%
\end{subequations}%
where the degeneracy factor $3(2j+1)=6|\kappa|$ of each single particle levels is considered. The quark numbers $N_i\ (i=u,d,s)$ are obtained by integrating the density $n_i(r)$ in coordinate space as
\begin{equation}
 N_i = \int 4\pi r^2 n_i(r) \mbox{d}r. \label{Eq:Ni}
\end{equation}
At a fixed total baryon number $A$, our calculation is then carried out by varying the chemical potential $\mu_\mathrm{b}$ so that $3A=N_u+N_d+N_s$. The total mass and free energy of the strangelet can finally be obtained as
\begin{eqnarray}
M &=& \sum_{i, n, \kappa} 6|\kappa| f_{in\kappa} \varepsilon_{in\kappa} - \int 12\pi r^2 n_\mathrm{b} V_V \mbox{d}r - E_\mathrm{C}, \label{Eq:M} \\
F &=& \sum_{i, n, \kappa} 6|\kappa| f_{in\kappa} \left\{\mu_i - T \ln\left[1+e^{-(\varepsilon_{in\kappa}-\mu_i)/T}  \right] \right\}  \nonumber \\
  &\mathrm{}& - \int 12\pi r^2 n_\mathrm{b} V_V \mbox{d}r  - E_\mathrm{C}, \label{Eq:F}
\end{eqnarray}
where the Coulomb energy is determined by
\begin{equation}
 E_\mathrm{C} = 2\pi \int r^2 n_\mathrm{ch}(r) e A_0(r) \mbox{d}r. \label{Eq:EC}
\end{equation}
The entropy is then fixed according to basic thermodynamic relation, i.e.,
\begin{equation}
 S = \frac{M-F}{T}. \label{Eq:S}
\end{equation}

At fixed baryon number $A$, temperature $T$, and parameters $C$ and $D$, the Dirac Eq.~\eqref{Eq:RDirac}, mean field potentials in Eqs.~\eqref{Eq:Vs} and \eqref{Eq:Vv}, Klein-Gordon Eq.~\eqref{Eq:K-G}, and density profiles in Eq.~\eqref{Eq:Density} are solved iteratively inside a box with the grid width less than 0.005 fm. Note that in our calculation we have introduced cutoffs on the quantum numbers $n\leq n_\mathrm{max}$ and $|\kappa|\leq \kappa_\mathrm{max}$, which are fixed by the criterion $6|\kappa| f_{in\kappa} \leq 10^{-6}$.

\subsection{\label{sec:Emission rates} Neutron and proton emission rates}
With the strangelet's properties fixed in Sec.~\ref{sec:the_SletSM}, the neutron and proton emission widths can then be estimated by examine the external neutron and proton gas densities that are in equilibrium with the strangelet, which are obtained with the usual formulae for nuclear reaction rates in the theory of nucleosynthesis~\cite{Bonche1984_NPA427-278, Zhu2014_PRC90-054316}, i.e.,
\begin{equation}
\Gamma_{p,n} = n_{p,n} \langle\sigma_{p,n}v_{p,n}\rangle. \label{Eq:emi}
\end{equation}
Here $\sigma_{p,n}$ is the neutron and proton capture cross sections of the strangelet, and $\langle\sigma_{p,n}v_{p,n}\rangle$ represents a statistical average over the states in the external gas with $v_{p,n}$ being the velocities of protons and neutrons.

The particle number densities of neutron and proton gases outside the hot strangelet can be derived as
\begin{equation}
n_{p,n} = \frac{1}{\pi^2} \int_0^\infty \left[1 + e^{(\sqrt{p^2 + m_{p,n}^2} - \mu_\mathrm{b})/T}\right]^{-1} p^2 \mbox{d}p,
\label{Eq:particle density}%
\end{equation}%
where the chemical potential $\mu_\mathrm{b}$ is fixed by Eq.~\eqref{Eq:bstb} as the strangelet is in thermodynamic equilibrium with the nucleon gas. Note that we have neglected the contributions of antinucleons as the temperatures considered here are relatively small.

The statistical average $\langle\sigma_{p,n}v_{p,n}\rangle$ can be calculated with
\begin{equation}
\langle\sigma_i v_i\rangle = \frac{\int_0^\infty  \sigma_i (\varepsilon_i) f(\varepsilon_i) v_i(\varepsilon_i) \sqrt{\varepsilon_i}\mbox{d}\varepsilon_i}
                                  {\int_0^\infty f(\varepsilon_i)\sqrt{\varepsilon} \mbox{d}\varepsilon_i},
\end{equation}
where $f(\varepsilon_i)$ represents the Fermi-Dirac distribution in Eq.~\eqref{Eq:F-D} with the discretized single particle energy $\varepsilon_{in\kappa}$ replaced by the continuum one $\varepsilon_i$ and chemical potential $\mu_i$ by $\mu_\mathrm{b}-m_i$. Note that the kinetic energy is connected to the velocity $v_i$ of nucleons with $\varepsilon_i = \sqrt{p_i^2 + m_i^2} - m_i\approx m_i v_i^2/2$, which gives
\begin{equation}
v_i(\varepsilon_i) = \sqrt{\frac{2\varepsilon_i}{m_i}}. \label{Eq:vi}
\end{equation}
Since neutron is an electrically neutral particle, we can simply adopt the geometrical cross section, i.e.,
\begin{equation}
\sigma_n = \pi R^2, \label{Eq:sigma n}
\end{equation}
where $R$ is the radius of the corresponding strangelet and we take its value at vanishing quark densities, i.e., boundary of the box. For the capture cross section of protons, the Coulomb interaction can not be neglected~\cite{Lai2021_RAA21-250, Wong1973_PRL31-766}. Then we adopt the Hill-Wheeler formula~\cite{Hill1953_PR089-1102} and assume a typical Coulomb barrier width $\omega_0 = 4$ MeV for nuclear reaction, i.e.,
\begin{equation}
\sigma_p(\varepsilon_p) = \frac{R^2\omega_0}{2\varepsilon_p}\ln\left\{1 + \exp\left[\frac{2\pi(\varepsilon_p-\varepsilon_C)}{\omega_0}\right]\right\}. \label{Eq:sigma p}
\end{equation}
The Coulomb barrier height is fixed at the Box boundary with $\varepsilon_C = {\alpha Z}/{R}$, where $\alpha$ = 1/137.036 is the fine structure constant and $Z$ the charge number carried by the strangelet.

\section{\label{sec:num}Results and discussions}
Based on the formulae presented in Sec.~\ref{sec:the_SletSM}, the properties of $\beta$-stable strangelets are then obtained at fixed baryon number $A$, temperature $T$, and parameter set $(C,\sqrt{D})$. Our previous investigations have revealed the properties of strangelets and $ud$QM nuggets at vanishing temperatures~\cite{Xia2018_PRD98-034031, Xia2019_AIPCP2127-020029, Xia2022_PRD106-034016}, which are sensitive to the strengths of confinement $D$ and perturbation $C$ interactions. In this work, we thus focus on the effects of finite temperature on the properties of strangelets.

\begin{figure}
\includegraphics[width=0.8\linewidth]{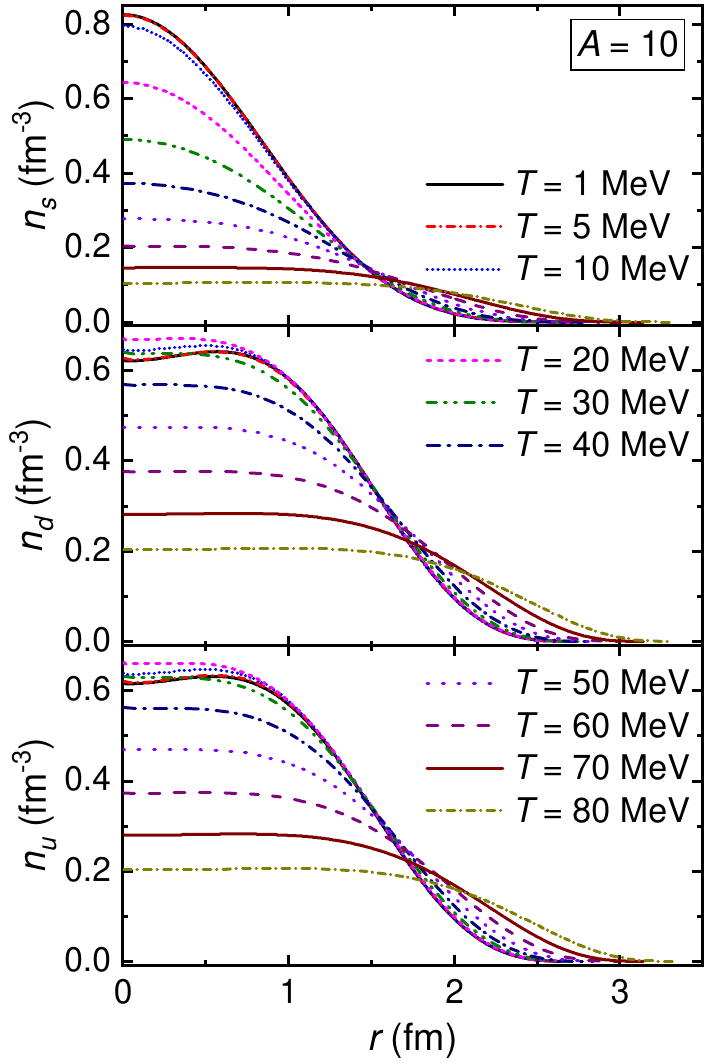}
\caption{\label{Fig:udsdens} Density profiles of $u$, $d$, $s$ quarks for strangelets at baryon number $A=10$ and various temperatures, where the parameter set $C=-0.5$ and $\sqrt{D}=180$ MeV is adopted.}
\end{figure}

\begin{figure}
\includegraphics[width=0.8\linewidth]{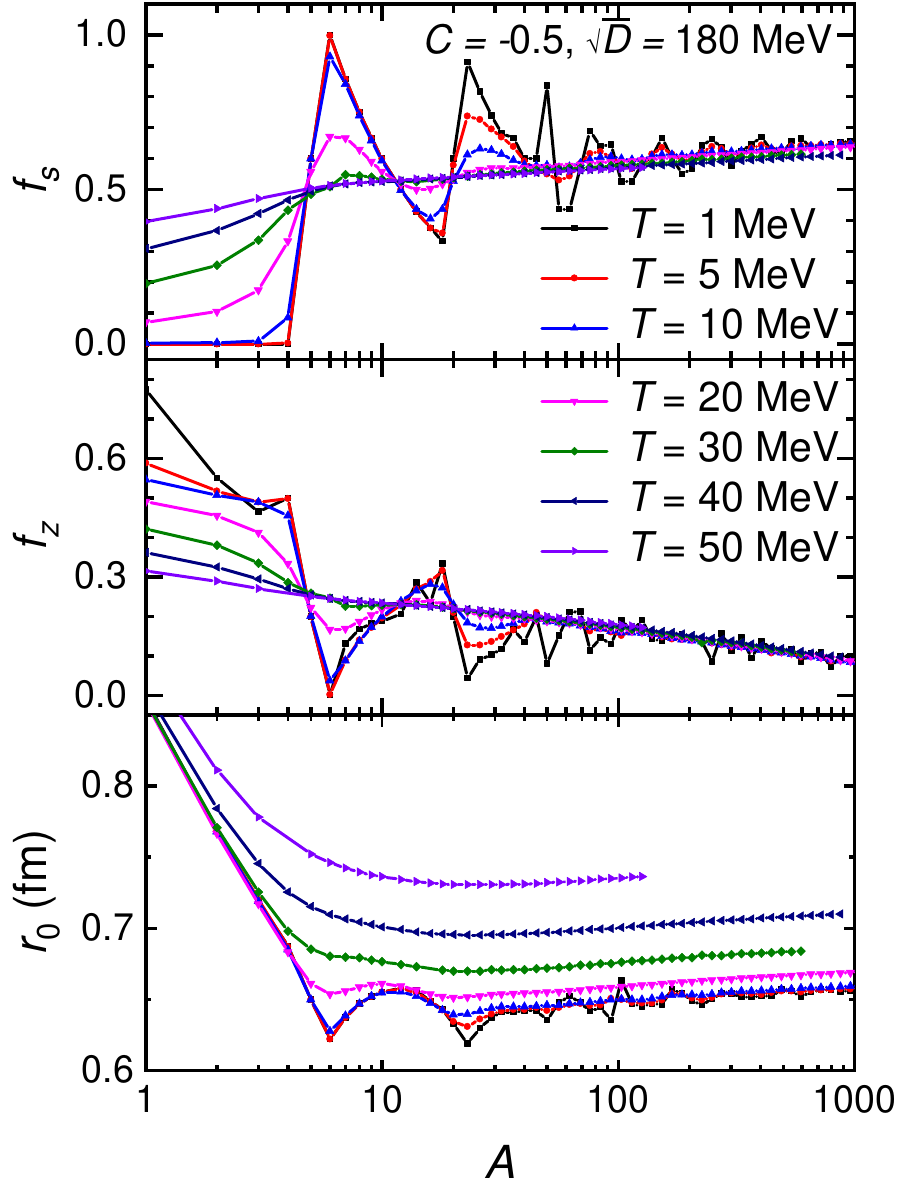}
\caption{\label{Fig:fsfzr0} Charge-to-mass ratio $f_z$, strangeness per baryon $f_s$, and ratio of root-mean-square radius to baryon number $r_0$ for $\beta$-stable strangelets obtained at various temperatures. }
\end{figure}

In Fig.~\ref{Fig:udsdens} we present the density profiles of $u$, $d$, $s$ quarks for strangelets at baryon number $A=10$ and temperatures $T = 1$-80 MeV, where the parameter set $C=-0.5$ and $\sqrt{D}=180$ MeV is adopted. As temperature increases, the densities of $u$, $d$, $s$ quarks decrease considerably at $T\gtrsim 20$ MeV. At $T\lesssim 20$ MeV, the shell effects play important roles, where the densities of $u$ and $d$ quarks are increasing with $T$ while that of $s$ quarks is decreasing. This is mainly attributed to the decreasing strangeness per baryon $f_s=N_s/A$ of the $\beta$-stable strangelet as $T$ increases, which is illustrated in Fig.~\ref{Fig:fsfzr0}. In fact, as indicated in Table.~\ref{table:prop}, the corresponding densities of SQM in the bulk limit fixed at vanishing pressures decrease with $T$ even at $T\lesssim 20$ MeV, which should be the case for strangelets at $A\gg 10$. Note that if we adopt parameter sets predicting small bulk densities, e.g., $C=0.7$ and $\sqrt{D}=129$ MeV or 140 MeV in Table.~\ref{table:prop}, the bulk density of $s$ quarks may increase with temperature. In the vicinity of quark-vacuum interface, the density distribution becomes more diffused as temperature increases. The variations in the density profiles of $u$ and $d$ quarks generally have similar trends as temperature increases, while the density of $s$ quarks varies more significantly.

\begin{figure}
\includegraphics[width=0.8\linewidth]{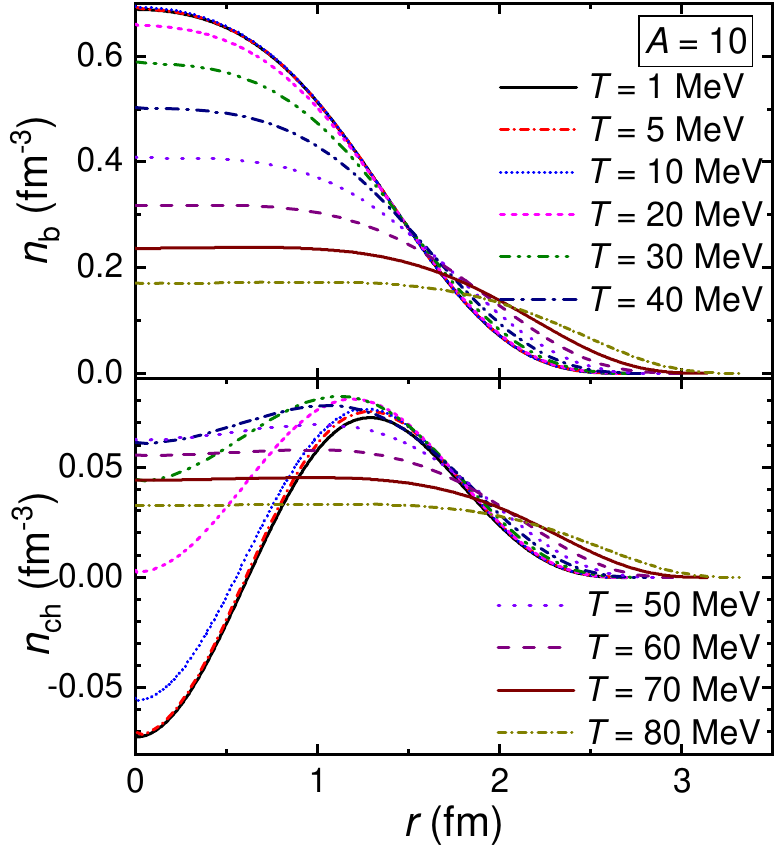}
\caption{\label{Fig:2mC-05A10} Same as Fig.~\ref{Fig:udsdens} but for the baryon and charge density profiles.}
\end{figure}

Figure~\ref{Fig:2mC-05A10} illustrates the charge $n_\mathrm{ch}=(2n_u-n_d-n_s)/3$ and baryon $n_\mathrm{b}=(n_u+n_d+n_s)/3$ density profiles of the strangelets corresponding to Fig.~\ref{Fig:udsdens}. As temperature increases, the baryon number density inside the strangelet decreases and becomes more diffused in the surface region. Similar trend is observed for baryon number density $n_0$ in the bulk limit as indicated in Table.~\ref{table:prop}, while the exact value is much smaller than that of the strangelets except for the case at $T=80$ MeV. At small temperatures, the charge density is negative at the center and increases as we move to the surface until reaching its peak at $r\approx 1.25$ fm, then decreases and finally tends to zero. As temperature increases, the variation of charge density becomes more smooth and positive, which is mainly attributed to the variations of the density profiles of $s$ quarks as indicated in Fig.~\ref{Fig:udsdens}.

\begin{figure}
\includegraphics[width=0.8\linewidth]{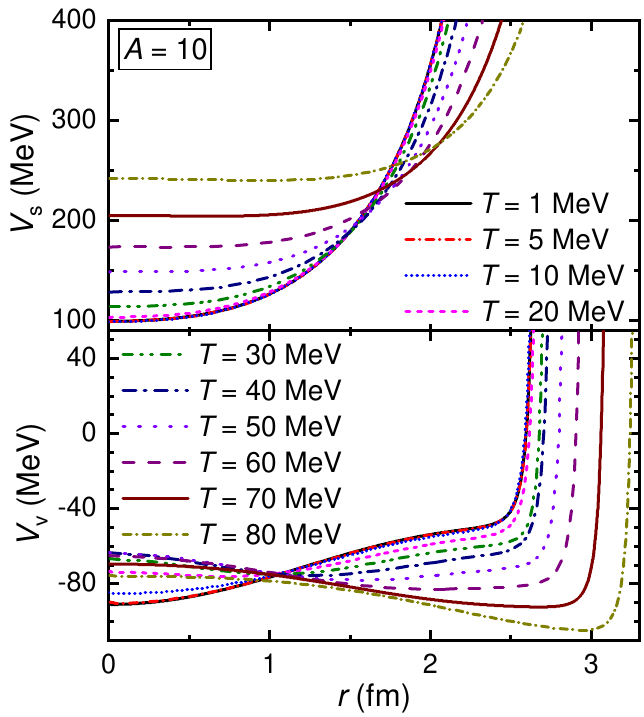}
\caption{\label{Fig:C-05V2A10} Scalar and vector potentials corresponding to the density profiles of the strangelets indicated in Figs.~\ref{Fig:udsdens} and \ref{Fig:2mC-05A10}.}
\end{figure}

Based on the density profiles indicated in Figs.~\ref{Fig:udsdens} and \ref{Fig:2mC-05A10}, the corresponding scalar and vector potentials are obtained with Eqs.~\eqref{Eq:Vs} and \eqref{Eq:Vv}, which are presented in Fig.~\ref{Fig:C-05V2A10}. As the effects of quark confinement are considered self-consistently in our mass scaling in Eq.~\eqref{eq:mI_CD}, as illustrated in Fig.~\ref{Fig:C-05V2A10}, the mean-field potentials become infinitely large in the vicinity of the quark-vacuum interface. As indicated in Figs.~\ref{Fig:udsdens} and \ref{Fig:2mC-05A10},  the densities of strangelets decrease with temperature and strangelets become larger in size. Consequently, the depth of scalar potential $V_s$ becomes shallower at higher temperatures, while the variation of the vector potential $V_v$ is smaller except that the position of infinite potential increases with the strangelet's size. Finally, at large enough temperatures, the densities become too small and the strangelet's size grows drastically. The mean fields then approach to the bulk limit, indicating the nonexistence of strangelets at larger temperatures.

\begin{figure}
\includegraphics[width=0.9\linewidth]{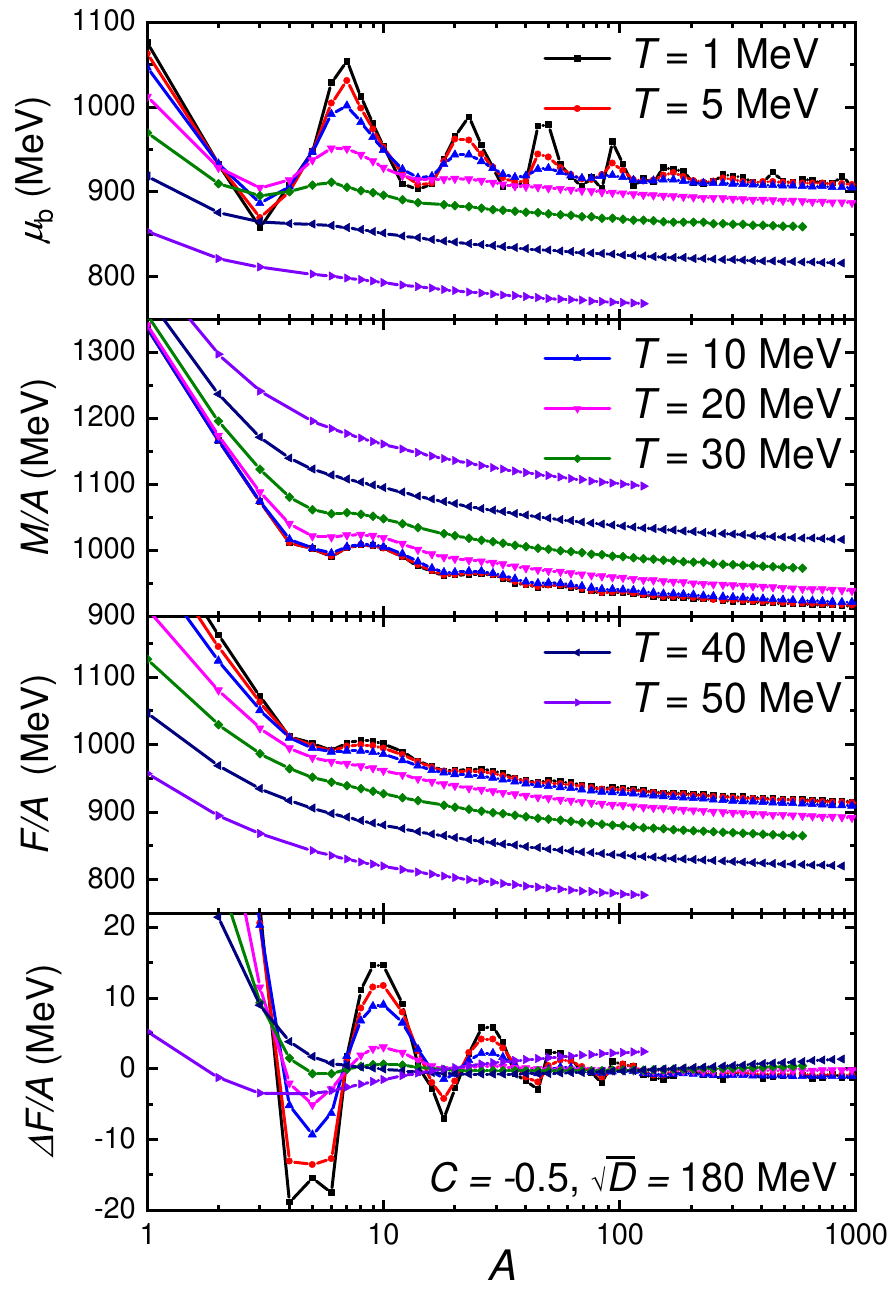}
\caption{\label{Fig:fpaepa} Chemical potential, energy and free energy per baryon of $\beta$-stable strangelets as functions of baryon number. The deviation in the free energy per baryon $\Delta F/A$ from the liquid-drop formula in Eq.~\eqref{Eq:fpa_LD} is presented in the lower panel as well.}
\end{figure}

In Fig.~\ref{Fig:fpaepa}, we present the chemical potential, energy and free energy per baryon for $\beta$-stable strangelets as functions of baryon number $A$ at different temperatures, where the parameter set $C=-0.5$ and $\sqrt{D}=180$ MeV is adopted. The chemical potential, energy and free energy per baryon are generally decreasing with $A$ and approaching to the bulk values indicated in Table.~\ref{table:prop}, i.e., $\mu_\mathrm{b}\rightarrow F_0/A$, $M/A\rightarrow (TS_0+F_0)/A$, and $F/A\rightarrow F_0/A$, which are fixed by treating quark matter at vanishing pressure while fulfilling both charge neutrality condition $n_\mathrm{ch}=0$ and $\beta$-stability condition $\mu_d=\mu_s$. According to the Fermi-Dirac distribution indicated in Eq.~\eqref{Eq:F-D}, quarks tend to occupy higher energy states as temperature increases, so that the energy per baryon increases with $T$, while the free energy per baryon and chemical potential decrease. This is also consistent with the trends in their bulk values in Table.~\ref{table:prop}. At a given temperature, the variation of strangelets' properties with respect to the adopted parameters $C$ and $D$ is consistent with our previous study at vanishing temperature~\cite{Xia2018_PRD98-034031}, where both the confinement and perturbation interactions destabilize the strangelets. For cold strangelets with small baryon numbers, the chemical potential, energy and free energy per baryon fluctuate with $A$, which is attributed to the shell effects~\cite{Xia2018_PRD98-034031}. Nevertheless, at larger temperatures, the shell effects are dampened so that $\mu_\mathrm{b}$, $M/A$ and $F/A$ varies smoothly with $A$. We should also mention that at small baryon numbers with $A\lesssim 10$, the center-of-mass correction and one-gluon-exchange interactions become important~\cite{Aerts1978_PRD17-260}, where the energy and free energy per baryon for $\beta$-stable strangelets are expected to be reduced. In the extreme case of $A=1$ and $T=0$, the (free) energy per baryon should correspond to the mass of nucleons~\cite{YOU2022_NPR39-302}.

\begin{figure}
\includegraphics[width=0.7\linewidth]{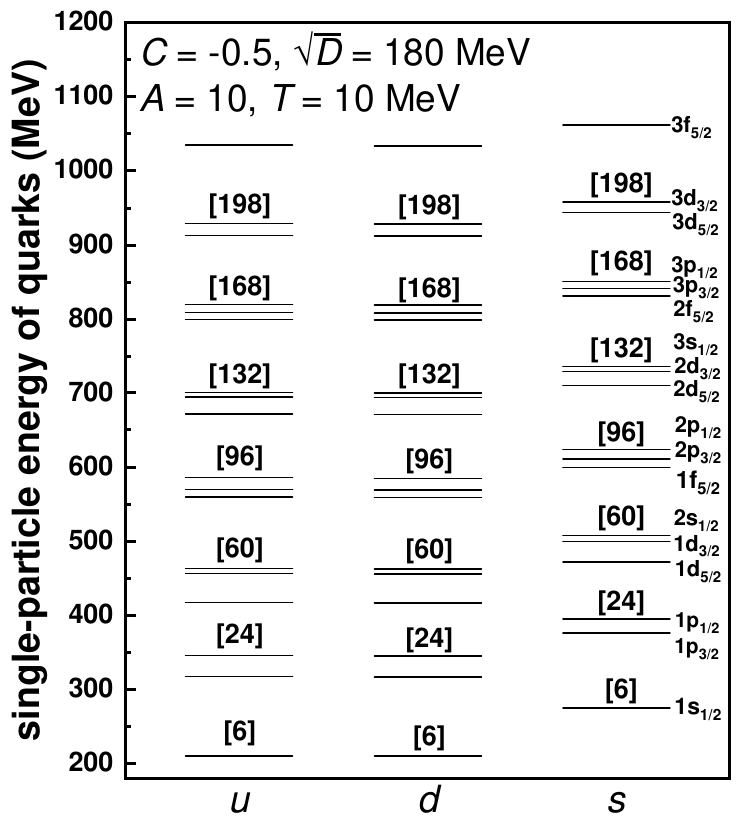}
\caption{\label{Fig:spl} Single-particle levels for $u$, $d$, $s$ quarks of the strangelet indicated in Figs.~\ref{Fig:udsdens}-\ref{Fig:C-05V2A10} at $T=10$ MeV. The magic numbers for quarks are indicated with the numbers inside the square brackets, where strangelets become more stable if the quark numbers $N_i$ take those values.}
\end{figure}

The free energy per baryon in Fig.~\ref{Fig:fpaepa} can be fitted by a liquid-drop type formula~\cite{Oertel2008_PRD77-074015}, i.e.,
\begin{equation}
\frac{F_\mathrm{LD}}{A} = \frac{F_0}{A} + \frac{\alpha_S}{A^{\frac{1}{3}}} + \frac{\alpha_C}{A^{\frac{2}{3}}}.
\label{Eq:fpa_LD}
\end{equation}
Here ${F_0}/{A}$ is the free energy per baryon in the bulk limit as indicated in Table.~\ref{table:prop}, while the fitted parameters $\alpha_S$ and $\alpha_C$ are presented as well. Note that during fitting we have subtracted the contribution of Coulomb energy $E_\mathrm{C}$ to better illustrate the effects of strong interaction on the interface effects.
The deviations of the free energy from the fitted values $\Delta F = F - E_\mathrm{C} - F_\mathrm{LD}$ are then presented in the bottom panel of Fig.~\ref{Fig:fpaepa}, where the shell effects can be identified easily. To better illustrate the shell structures, in Fig.~\ref{Fig:spl} we present the single-particle levels for $u$, $d$, $s$ quarks in the strangelet at $A=10$ and $T=10$ MeV, where the parameter set $C=-0.5$ and $\sqrt{D}=180$ MeV is adopted. Various magic numbers (6, 24, 60, 96, ...) with large shell gaps can be identified for $u$, $d$, $s$ quarks, which resembles the magic numbers (2, 8, 20, 28, ...) of finite nuclei. Note that the magic numbers at $N_i>24$ may be altered for larger strangelets or if we adopt different parameter sets~\cite{Xia2019_AIPCP2127-020029}, while the magic numbers 6 and 24 for quarks seem robust despite the variations in the adopted parameters. Consequently, as indicated in Fig.~\ref{Fig:fpaepa}, the strangelets at $A=4$, 6, and 18 are found to be more stable than others, which correspond to the quark numbers ($N_u$, $N_d$, $N_s$): (6, 6, 0), (6, 6, 6), and (24, 24, 6). As baryon number $A$ increases, the shell corrections to the energy and free energy per baryon eventually become insignificant. Similarly, if we adopt larger temperatures, quarks start to occupy more single-particle states above the Fermi energy as illustrated in Fig.~\ref{Fig:spl}, which smears out the fluctuations caused by the large shell gaps, i.e., shell dampening.

\begin{figure}
\includegraphics[width=0.8\linewidth]{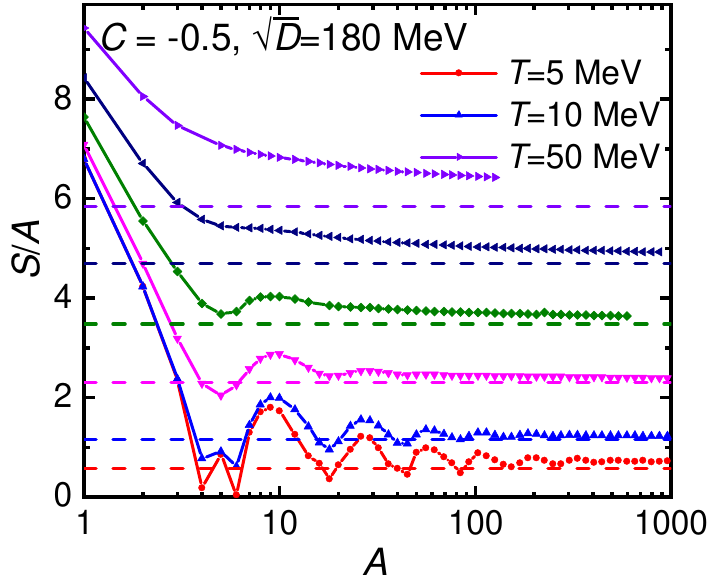}
\caption{\label{Fig:Spa-T} Entropy per baryon as functions of baryon number for $\beta$-stable strangelets obtained at various temperatures. The dashed horizontal lines correspond to the bulk values $S_0/A$ indicated in Table.~\ref{table:prop}.}
\end{figure}

Based on the energy and free energy per baryon indicated in Fig.~\ref{Fig:fpaepa}, the entropy per baryon can be obtained with Eq.~\eqref{Eq:S}, which are presented in Fig.~\ref{Fig:Spa-T} as functions of baryon number for $\beta$-stable strangelets. It is found that $S/A$ generally decreases with $A$ and approaches to its bulk limit $S_0/A$ in Table.~\ref{table:prop}, which is indicated with the dashed horizontal lines. Evidently, the entropy per baryon approaches faster to its bulk limit than energy and free energy per baryon, where at $A\geq4$ the value of $S/A$ is already close to $S_0/A$. Slight fluctuations caused by shell effects is also identified while the amplitude decreases with temperature due to shell dampening. The entropy per baryon increases quickly with $T$, which is also the case for the bulk limit indicated Table.~\ref{table:prop}. The dependence of $S/A$ on the adopted parameter sets were investigated in Ref.~\cite{Chen2022_PRD105-014011} and can also be identified in Table~\ref{table:prop}, where perturbative interaction reduces the entropy per baryon and confinement interaction does the opposite.

The strangeness per baryon, charge-to-mass ratio, and ratio of root-mean-square radius to baryon number can be obtained with
\begin{eqnarray}
  f_s &=& N_s/A,  \\
  f_z &=& Z/A = \left(2N_u - N_d -N_s\right)/3A, \\
  r_0 &=& {\left(\int 4\pi r^4 n_\mathrm{b} \mbox{d}r\right)}^{1/2}/A^{5/6},
\end{eqnarray}
which are then presented in Fig.~\ref{Fig:fsfzr0}. Generally speaking, the charge-to-mass ratio and ratio of root-mean-square radius to baryon number decrease with $A$ and eventually approach to their bulk values ($f_z = 0$ and $r_0 \approx 0.48/n_0^{1/3}$), while the strangeness per baryon increases with $A$ and approaches to its bulk value ($f_s \approx 0.7$) as indicated in Table.~\ref{table:prop}. Due to shell effects, the strangeness per baryon and charge-to-mass ratio fluctuates with the sequential occupation of lowest energy levels at small $T$ and eventually become smooth at larger $A$ or $T$. Note that the shells have opposite effects on $f_z$ and $f_s$, where $f_z$ suddenly drops if $s$-quark shell is occupied with a sudden increase of $f_s$. This will also alter the radii of strangelets, where a sudden decrease in $r_0$ is observed if $f_s$ increases. As temperature increases, the shell effects are dampened with $f_z$, $f_s$ and $r_0$ vary more smoothly with $A$. For the parameter sets adopted here with $C=-0.5$ and $\sqrt{D}=180$ MeV, $f_z$ and $f_s$ are generally insensitive to the temperature effects other than shell dampening. This is very different from the cases predicting small saturation densities, where $f_s$ increases with $T$, e.g., adopting the parameters $C=0.4$, 0.7 and $\sqrt{D}=129$, 140 MeV as indicated in Table.~\ref{table:prop}. Meanwhile, as illustrated in Figs.~\ref{Fig:udsdens} and \ref{Fig:2mC-05A10}, the density profiles in strangelets become more dilute as temperature increases, leading to larger $r_0$ in Fig.~\ref{Fig:fsfzr0}.

\begin{figure}
\includegraphics[width=0.8\linewidth]{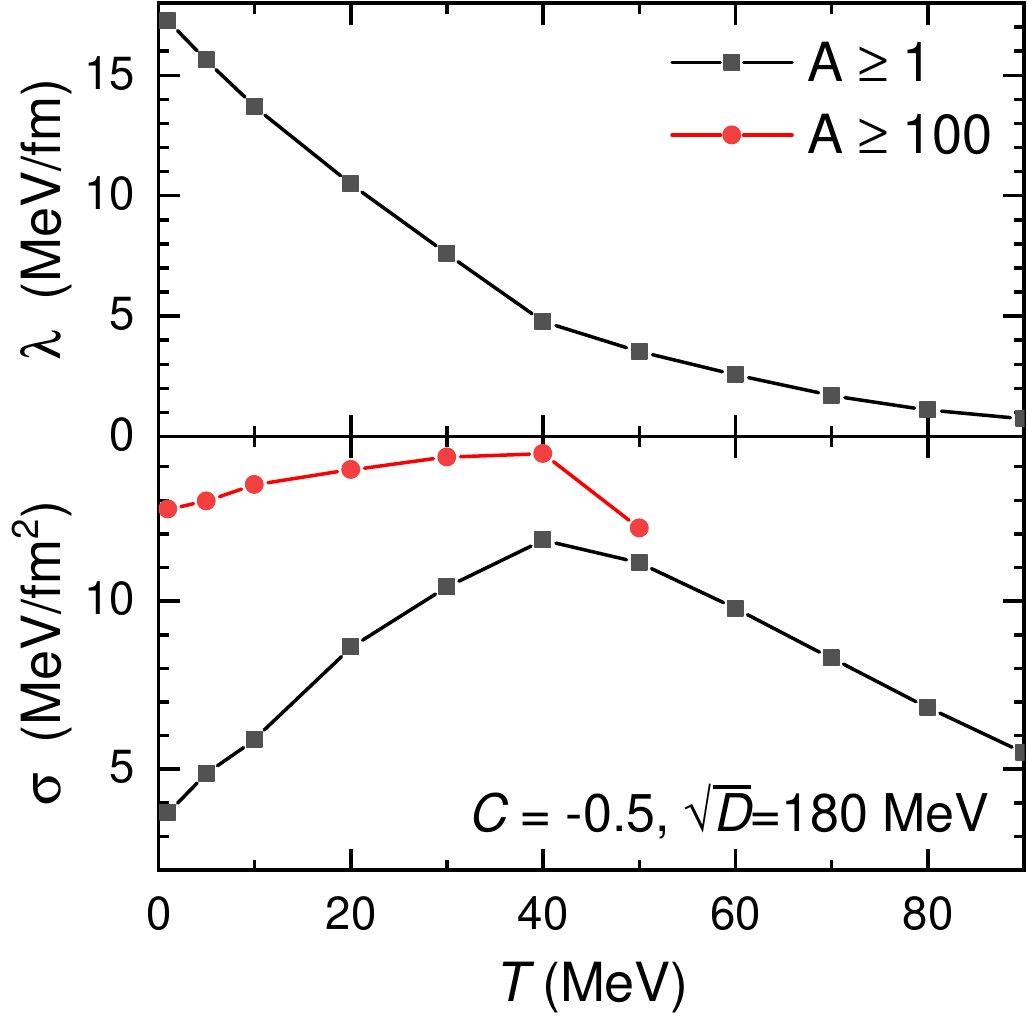}
\caption{\label{Fig:Sur-Cur-T} The obtained surface tension $\sigma$ and curvature term $\lambda$ of SQM as functions of temperature, which are fixed by Eqs.~\eqref{Eq:as} and \eqref{Eq:ac}. Two fitting procedures are adopted to fix the parameter $\alpha_S$, i.e., using the free energy per baryon of strangelets at $A\geq1$ and $A\geq100$.}
\end{figure}

Using the fitted parameters $\alpha_S$ and $\alpha_C$ of the liquid-drop type formula in Eq.~\eqref{Eq:fpa_LD}, the surface tension and curvature term of SQM can then be extracted with
\begin{eqnarray}
\sigma &=& \alpha_S\left(\frac{n_0^2}{36\pi}\right)^{1/3}, \label{Eq:as} \\
\lambda &=& \alpha_C\left(\frac{n_0}{384\pi^2}\right)^{1/3}, \label{Eq:ac}
\end{eqnarray}
which are presented in Table~\ref{table:prop}. To better illustrate the variations of surface tension and curvature term of SQM with respect to temperature, in Fig.~\ref{Fig:Sur-Cur-T} we plot the obtained values for $\sigma$ and $\lambda$ as functions of temperature, where the parameters $\alpha_S$ and $\alpha_C$ are fixed employing all the numerical results in Fig.~\ref{Fig:fpaepa}. As indicated by the black squares, the curvature term $\lambda$ decreases monotonically with temperature, while the surface tension increases with $T$ until reaching its peak at $T \approx 40$ MeV. Such type of behavior is also observed if we adopt other parameter sets as indicated in Table~\ref{table:prop}. Nevertheless, the increment of the surface tension at $T \lesssim 40$ MeV is exactly the opposite of what we have expected, where previous investigations typically suggest that $\sigma$ decreases with $T$, e.g., those in Refs.~\cite{Pinto2012_PRC86-025203, Mintz2013_PRD87-036004, Ke2014_PRD89-074041}. The reason for such type of behaviors in $\sigma$ is mainly attributed to the shell effects, where as temperature increases the free energy per baryon increase quickly for small strangelets due to shell dampening. In fact, to show this more explicitly, we have refitted the parameter $\alpha_S$ using only the numerical results of large strangelets with $A\geq100$ and kept $\alpha_C$ unchanged, where the obtained surface tension is plotted as red solid circles in Fig.~\ref{Fig:Sur-Cur-T}. Evidently, due to the diminished shell effects in large strangelets, the surface tension increases slightly and then drops quickly at $T \gtrsim 40$ MeV. In this work, we are more interested in describing the properties of all strangelets that may be created in binary compact star mergers and heavy-ion collisions, thus the numerical results for small strangelets are included as well during fitting, where the corresponding values are presented in Table~\ref{table:prop}.

\begin{table*}
\caption{\label{table:prop} Bulk properties of SQM at zero external pressure, interface properties obtained using the fitted liquid-drop parameters, and emission rates per unit surface area for large strangelets obtained at various temperatures and parameter sets.}
\begin{tabular}{ccc|cccc|cccc|cc} \hline \hline
\multicolumn{3}{c|}{Parameters} & \multicolumn{4}{c|}{Bulk properties}
& \multicolumn{4}{|c}{Interface properties} & \multicolumn{2}{|c}{Emission rates}  \\ \hline
$C$ & $\sqrt{D}$ & T &    $n_0$ & $f_s$    & $F_0/A$ &  $S_0/A$   & $\alpha_S$  & $\alpha_C$   & $\sigma$   & $\lambda$  & $\Gamma_n/4\pi R^2$ & $\Gamma_p/4\pi R^2$  \\
 & MeV & MeV & fm${}^{-3}$   & fm${}^{-3}$ &  MeV &   &  MeV & MeV & MeV/fm${}^2$ & MeV/fm & MeV/fm${}^2$ & MeV/fm${}^2$  \\ \hline
$-0.5$& 180 & 1  &  0.38  & 0.70 &  902  &  0.11  &  34  & 371 & 3.7  & 17.3 & $2.0\times10^{-19}$  & $7.9\times10^{-30}$  \\
$-0.5$& 180 & 5  &  0.38  & 0.70 &  900  &  0.57  &  45  & 337 & 4.9  & 15.6 & $1.8\times10^{-5}$   & $1.3\times10^{-7}$   \\
$-0.5$& 180 & 10 &  0.37  & 0.70 &  895  &  1.15  &  55  & 298 & 5.9  & 13.7 & 0.0013               & $1.1\times10^{-4}$   \\
$-0.5$& 180 & 20 &  0.36  & 0.69 &  877  &  2.30  &  82  & 230 & 8.6  & 10.5 & 0.022                & 0.0063               \\
$-0.5$& 180 & 30 &  0.34  & 0.69 &  846  &  3.47  &  103 & 170 & 10.4 & 7.60 & 0.051                & 0.022                \\
$-0.5$& 180 & 40 &  0.31  & 0.68 &  802  &  4.70  &  125 & 110 & 11.8 & 4.77 & 0.062                & 0.033                \\
$-0.5$& 180 & 50 &  0.29  & 0.68 &  746  &  5.84  &  123 & 83  & 11.2 & 3.53 & 0.073                & 0.043                \\
$-0.5$& 180 & 60 &  0.26  & 0.69 &  676  &  7.07  &  116 & 63  & 9.8  & 4.75 & 0.064                & 0.041                \\
$-0.5$& 180 & 70 &  0.22  & 0.69 &  589  &  8.47  &  110 & 44  & 8.3  & 1.69 & 0.044                & 0.030                \\
$-0.5$& 180 & 80 &  0.19  & 0.70 &  488  &  9.85  &  100 & 30  & 6.8  & 1.10 & 0.031                & 0.023                \\
$-0.5$& 180 & 90 &  0.16  & 0.71 &  367  &  11.5  &  90  & 22  & 5.5  & 0.75 & 0.021                & 0.016                \\ \hline
 0    & 156 & 1  &  0.24  & 0.52 &  913  &  0.15  &  87  & 178 & 6.9  & 7.09 & $8.9\times10^{-16}$  & $1.5\times10^{-28}$  \\
 0    & 156 & 5  &  0.24  & 0.52 &  911  &  0.74  &  90  & 159 & 7.2  & 6.34 & $9.4\times10^{-5}$   & $2.3\times10^{-7}$   \\
 0    & 156 & 10 &  0.24  & 0.52 &  905  &  1.47  &  102 & 117 & 8.1  & 4.67 & 0.0049               & $2.5\times10^{-4}$   \\
 0    & 156 & 20 &  0.22  & 0.51 &  881  &  2.98  &  113 & 86  & 8.5  & 3.33 & 0.026                & 0.0057               \\
 0    & 156 & 30 &  0.20  & 0.51 &  841  &  4.47  &  121 & 57  & 8.6  & 2.12 & 0.043                & 0.016                \\ \hline
 0.4  & 129 & 1  &  0.11  & 0.18 &  852  &  0.21  &  57  & 143 & 2.7  & 4.38 & $9.8\times10^{-44}$  & $1.4\times10^{-57}$  \\
 0.4  & 129 & 5  &  0.11  & 0.19 &  849  &  1.05  &  53  & 156 & 2.5  & 4.80 & $2.1\times10^{-10}$  & $3.2\times10^{-13}$  \\
 0.4  & 129 & 10 &  0.11  & 0.21 &  841  &  2.05  &  66  & 115 & 3.1  & 3.54 & $5.7\times10^{-6}$   & $2.4\times10^{-7}$   \\
 0.4  & 129 & 20 &  0.10  & 0.26 &  808  &  4.04  &  83  & 75  & 3.7  & 2.22 & $6.6\times10^{-4}$   & $1.5\times10^{-4}$   \\
 0.4  & 129 & 30 &  0.09  & 0.33 &  755  &  5.95  &  91  & 35  & 3.8  & 1.01 & 0.0028               & 0.0011               \\ \hline
 0.7  & 129 & 1  &  0.10  & 0.06 &  920  &  0.23  &  44  & 183 & 2.0  & 5.43 & $3.7\times10^{-12}$  & $5.0\times10^{-26}$  \\
 0.7  & 129 & 5  &  0.10  & 0.08 &  917  &  1.12  &  57  & 135 & 2.5  & 4.02 & $3.7\times10^{-4}$   & $6.0\times10^{-7}$   \\
 0.7  & 129 & 10 &  0.10  & 0.12 &  908  &  2.19  &  65  & 114 & 2.9  & 3.40 & 0.0069               & $3.0\times10^{-4}$   \\
 0.7  & 129 & 20 &  0.09  & 0.19 &  872  &  4.36  &  83  & 68  & 3.4  & 1.95 & 0.019                & 0.0044               \\
 0.7  & 129 & 30 &  0.07  & 0.25 &  814  &  6.74  &  88  & 39  & 3.1  & 1.04 & 0.013                & 0.0052               \\ \hline
 0.7  & 140 & 1  &  0.13  & 0.13 &  997  &  0.22  &  52  & 192 & 2.8  & 6.24 & 2.0                  & 0.31                 \\
 0.7  & 140 & 5  &  0.13  & 0.14 &  994  &  1.06  &  62  & 157 & 3.3  & 5.09 & 1.9                  & 0.32                 \\
 0.7  & 140 & 10 &  0.13  & 0.18 &  985  &  2.07  &  75  & 118 & 4.0  & 3.85 & 1.7                  & 0.33                 \\
 0.7  & 140 & 20 &  0.12  & 0.24 &  952  &  4.05  &  93  & 73  & 4.7  & 2.30 & 0.78                 & 0.22                 \\
 0.7  & 140 & 30 &  0.10  & 0.29 &  898  &  6.12  &  99  & 44  & 4.4  & 1.32 & 0.25                 & 0.10                 \\
\hline
\end{tabular}
\end{table*}

\begin{figure}
\includegraphics[width=0.9\linewidth]{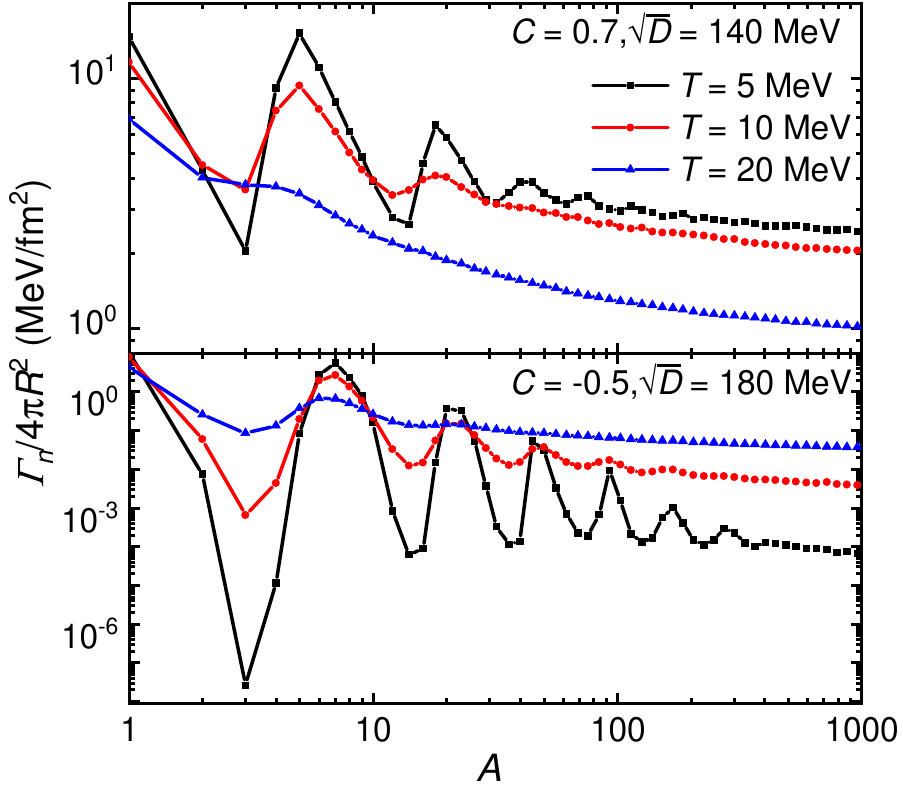}
\caption{\label{Fig:emi-T} Neutron emission rates for $\beta$-stable strangelets as functions of baryon number, which are fixed by Eq.~\eqref{Eq:emi}.}
\end{figure}

Based on the obtained chemical potentials for $\beta$-stable strangelets, the proton and neutron emission rates can then be fixed with Eq.~\eqref{Eq:emi}. As an example, in Fig.~\ref{Fig:emi-T} we present the neutron emission rates for $\beta$-stable strangelets adopting two different parameter sets, where $C=-0.5$ and $\sqrt{D}=180$ MeV predicts stable strangelets while $C=0.7$ and $\sqrt{D}=140$ MeV predicts unstable ones. Consequently, the strangelets obtained with $C=0.7$ and $\sqrt{D}=140$ MeV are unstable against neutron emission with large decay rates, which are decreasing with temperature. This is essentially different from the cases obtained with $C=-0.5$ and $\sqrt{D}=180$ MeV, where the neutron emission rates are small and increasing with $T$ at $T\lesssim 50$ MeV. The reason for such variation is mainly due to the evolution of chemical potential, where for unstable strangelets $\mu_\mathrm{b}>m_n$ and $\mu_\mathrm{b}$ eventually decreases as temperature increases. Due to shell effects, the chemical potential fluctuates with $A$ so that the emission rates also fluctuate, leading to extreme stable strangelets that might function as waiting points during the decay of large strangelets and finally persist in binary compact star mergers or heavy-ion collisions. At larger temperatures, such fluctuations in nucleon emission rates vanish due to shell dampening. In general, the neutron or proton emission rates per unit surface area of strangelets decrease with $A$ and eventually approach to their bulk values indicated in Table~\ref{table:prop}, which are obtained using the bulk chemical potential $\mu_\mathrm{b}=F_0/A$ and Coulomb barrier $\varepsilon_C = \mu_d-\mu_s$ according to the properties of SQM at zero external pressure. Note that the neutron or proton emission rates at $A\lesssim 10$ are almost the same while at larger $A$ the proton emission rates decrease faster than that of neutron, which finally approach to the bulk limit illustrated in Table~\ref{table:prop}. At larger temperatures, the neutron and proton emission rates per unit area may decrease further, which is attributed to the fast decrease of the chemical potential.

\section{\label{sec:con}Conclusion}
We study the properties of strangelets at finite temperature $T$ in an equivparticle model with both the linear confinement and leading-order perturbative interactions accounted for using density-dependent quark masses. The energy, free energy, entropy, charge-to-mass ratio, strangeness per baryon, and ratio of root-mean-square radius to baryon number for $\beta$-stable strangelets are obtained with various parameter sets. Adopting mean-field and no-sea approximation, the Dirac equations for quarks inside a strangelet are solved iteratively with the mean fields fixed by the corresponding density profiles. Various magic numbers (6, 24, 60, 96, ...) with large shell gaps are identified for $u$, $d$, $s$ quarks, where strangelets at baryon number $A=4$, 6, and 18 are found to be more stable than others. As temperature increases, the shell corrections diminish and eventually become insignificant as higher energy states are being occupied by quarks, i.e., shell dampening. Consequently, instead of decreasing with temperature, the surface tension extracted from a liquid-drop formula fitted to the free energies of strangelets at $A\geq 1$ increases with $T$ until reaching its peak at $T\approx 20$-40 MeV with vanishing shell corrections, while the curvature term decreases with $T$ monotonically. Based on the obtained properties of strangelets, the neutron and proton emission rates can then be fixed according to the usual formulae for nuclear reaction rates in the theory of nucleosynthesis, where the external nucleon gas densities in equilibrium with the strangelets and their capture cross sections are adopted. The obtained emission rates for stable strangelets are generally increasing with $T$ until reaching their peaks at $T\approx 20$-50 MeV, while for unstable ones the emission rates decrease monotonically with $T$. In fact, the emission rates are deeply connected to the chemical potential of strangelets, which fluctuates due to shell corrections so there exist strangelets that are extremely stable against neutron or proton emissions. Consequently, those strangelets might function as waiting points during the decay of large strangelets and finally persist in binary compact star mergers and heavy-ion collisions, which may potentially be observed.

\begin{acknowledgments}
The authors would like to thank Prof. Gentaro Watanabe and Prof. Nobutoshi Yasutake for fruitful discussions. This work was supported by the National Natural Science Foundation of China (Grant Nos. 12275234 and 12342027) and the National SKA Program of China (Grant Nos. 2020SKA0120300 and 2020SKA0120100).
\end{acknowledgments}


%

\end{document}